\documentclass[11pt,letterpaper]{JHEP3}
\usepackage{graphics}
\usepackage{epsfig}
\usepackage{subfigure}
\usepackage{mathrsfs}
\usepackage{amsmath}
\usepackage{amsfonts}
\usepackage{amssymb}
\usepackage{graphicx}
\usepackage{longtable}
\usepackage{rotating}
\usepackage{multirow}

\title{Phase structure of the Born-Infeld-anti-de Sitter black holes probed by non-local observables}
\author{Xiao-Xiong Zeng$^{1,2}$, Xian-Ming Liu$^{3,4}$, Li-Fang Li$^{5}$\\
$^{1}$ School of Material Science and Engineering, Chongqing Jiaotong University, Chongqing\\
        ~400074, China\\
$^{2}$ Institute of Theoretical Physics,
        Chinese Academy of Sciences,
        Beijing 100190, China\\
$^{3}$ Center for Theoretical Physics, Massachusetts Institute of Technology,
        Cambridge, \\
        ~MA 02139,USA\\
$^{4}$ Center for Theoretical Physics, School of Sciences, Hubei University for\\ ~
      Nationalities, Enshi, Hubei  445000, China\\
$^{5}$ State Key Laboratory of Space Weather, Center for Space Science and Applied Research,\\
        ~ Chinese Academy of Sciences, Beijing 100190, China\\
 \email{xxzeng@itp.ac.cn;  liuxianming1980@163.com; lilf@itp.ac.cn}
}
 \abstract{
With the non-local observables such as  two point correlation function and holographic entanglement entropy, we probe the phase structure of the  Born-Infeld-anti-de Sitter black holes. We find for the case $bQ>0.5$, the phase structure is similar to that of the Reissner-Nordstr\"{o}m-AdS  black hole, namely the black hole undergoes a Hawking-Page phase transition, a first order phase transition, and a second order phase transition. While for the case $bQ<0.5$, we find there is a new branch for the  infinitesimally  small black hole  so that a pseudo phase transition emerges besides the original first order phase transition. For the first order phase transition and the pseudo phase transition, the equal area law is checked, and for the second order phase transition, the critical exponent of the analogous heat capacity is obtained in the neighborhood of the critical points. All the results show that the phase structure of the  non-local observables is the same as that of the thermal entropy regardless of the size of the boundary region in the field theory.
}

\preprint{}
\begin{document}
\bibliographystyle{ieeetr}

\section{Introduction}

Investigation on phase transition of an AdS space time  has attracted attention of many theoretical physicists recently.  The main motivation maybe steps from the existence of the   AdS/CFT correspondence \cite{ads1,ads2,ads3}, which relates a black hole in the AdS space time  to a thermal system without gravity. For in this case, some interesting but intractable phenomena in strongly coupled system become to be tractable easily in the bulk.  To probe these  fascinating phenomena in field theory, one should employ some non-local observables  such as two point correlation function, Wilson loop, and holographic entanglement entropy, which are dual to the geodesic length, minimal area surface, and minimal surface area  in the bulk individually. It has been shown that these observables can probe  the non-equilibrium thermalization behavior \cite{
Balasubramanian1,Balasubramanian2,GS,CK, Zeng2013, Zeng2014,Zeng20151,Liuh,Zhangs,Buchel, Craps}, superconducting phase transition \cite{
Superconductors1, Superconductors2, Superconductors3, Superconductors4, Superconductors5, Superconductors7,
Superconductors8,Ling}, and cosmological singularity \cite{Engelhardt,Engelhardt1}.

In this paper, we intend to use the non-local observables to probe the phase structure of the  Born-Infeld-anti-de Sitter black holes.  Usually,  phase structure of a  black hole is understood  from the viewpoint of thermodynamics.
For an uncharged AdS black hole,
it has been found that there is a phase transition between the  thermal gas in AdS space and Schwarzschild AdS black holes  \cite{Hawking4},  which was interpreted as the confinement/deconfinement phase
transition in the dual gauge field theory later \cite{Witten1}.  As the charge is endowed with,  the  AdS black hole  will undergo  a Van der Waals-like phase transition  before
it reaches the stable state in the entropy-temperature plane \cite{Chamblin}.
Specifically speaking, there exists a critical charge, and  for the case that the charge of the black hole is smaller than the critical charge, the black hole
 undertakes a first order phase transition.   As the charge increases to the critical charge, the
 phase transition is second order, while  the charge exceeds
the critical charge, there is not phase transition and the black hole is always stable.
The Van der Waals-like phase transition can be observed in many  circumstances. In  \cite{Cai}, it was found that a 5-dimensional neutral Gauss-Bonnet
black hole demonstrates the Van der Waals-like phase transition in the $T-\alpha$ plane, where $T$ is the Hawking temperature and $\alpha$ is the Gauss-Bonnet coupling parameter. In  \cite{Niu}, the Van
der Waals-like phase transition was also observed in the $Q-\Phi$ plane,  where $Q$ is electric charge and $\Phi$ is the chemical potential. Treating the negative cosmological constant  as the pressure $P$ and its conjugate as
the thermodynamical volume $V$, the Van der Waals-like phase transition has also been observed
in the $P-V$ plane recently\cite{Kastor,Kubiznak,
Xu, Caipv, Hendi, Hennigar, Wei1, Mo,Spallucci}.

Very recently, with the entanglement entropy as a probe, \cite{Johnson} investigated the phase structure of the  Reissner-Nordstr\"{o}m-AdS  black hole and found that there was also a
Van der Waals-like phase transition in the entanglement entropy-temperature plane \footnote{ In fact, the Van der Waals phase transition has also been observed  for a 
zero temperature system  in the framework of holography \cite{Bigazzi}.}. They also obtained the critical exponent of the heat capacity  for the second order phase transition in the neighborhood of the critical points.
 To further confirm the similarity of the phase structure between the thermal entropy and entanglement entropy,
\cite{Nguyen} checked the equal area law later and found that it held for the first order phase transition in the entanglement entropy-temperature plane.
Now \cite{Johnson} has been generalized to the extended space time \cite{Caceres}, massive gravity \cite{zeng2016} , as well as Weyl  gravity \cite{Dey}, and all the results showed that the entanglement entropy exhibited the same phase structure as that of the thermal entropy.

In this paper, besides the entanglement entropy, we will  employ the two point correlation function to probe the phase structure of the black holes. We choose the  Born-Infeld-anti-de Sitter black holes as the gravity model,  which   is a solution of the Einstein-Born-Infeld action. There have been many works to study  phase structure of the   Born-Infeld-anti-de Sitter black holes  \cite{Myung,Lala,Hendibi,Zouni}. The results showed that both the Born-Infeld parameter $b$ and charge $Q$ affect the phase structure, and to  assure the existence of a non-extremal  black hole, one should
 impose the condition $bQ>0.5$. In this paper, we find for the case  $bQ>0.5$, the phase structure of the black hole is similar to that of the Reissner-Nordstr\"{o}m-AdS  black hole in the entropy-temperature plane. While for the case $bQ<0.5$, there is a novel phase structure, which has not been observed previously. Specially speaking,  a new branch emerges compared with that of the Reissner-Nordstr\"{o}m-AdS  black hole so that there are two unstable regions and two phase transition temperature correspondingly.
 All these phase structure are probed by the nonlocal observables  such as two point correlation function as well as holographic entanglement entropy, and phase structure of the nonlocal observables  are found to be the same as that of the thermal entropy.

Our paper is outlined as follows. In the next section, we will review the thermodynamic properties of the Born-Infeld-anti-de Sitter  black hole firstly, and then study its  phase structure in $T-S$ plane in a fixed charge ensemble. In Section~\ref{Nonlocal_observables}, we will employ the two point correlation function and  holographic entanglement entropy to probe the phase structure of the Born-Infeld-anti-de Sitter  black hole. In each subsection, the  equal area law is checked and the critical exponent of the heat capacity is obtained.
The last section is devoted to discussions and conclusions.

\section{Thermodynamic phase transition of the Born-Infeld-anti-de Sitter  black hole}
\subsection{Review of the Born-Infeld-anti-de Sitter  black hole}
\label{quintessence_Vaidya_AdS}
The 4-dimensional Born-Infeld AdS black hole  is a solution of the following action \cite{Rasheed}
\begin{equation}
S=\int d^4 x \sqrt{-g}[\frac{R-3 \Lambda }{16 \pi G} +\frac{b^2}{4 \pi G}(1-\sqrt{1+\frac{2F}{b^2}})],
\end {equation}
in which $F=\frac{1}{4}F_{\mu\nu}F^{\mu\nu}$, $R$ is scalar curvature, $G$ is the gravitational constant,
 $ \Lambda$ is the cosmological constant that relates  to the AdS radius as $ \Lambda=-3/l^2$, and $b$ is the Born-Infeld parameter which  relates to the string tension $\alpha^{\prime}$ as $b = 1/(2 \pi \alpha^{\prime})$.
Explicitly, the solutions of the Born-Infeld AdS black hole can be written as \cite{Caibiads,Deybi}
\begin{equation}
 ds^{2}=-f(r)dt^{2}+f^{-1}(r)dr^{2}+r^{2}(d\theta^2+\sin^2\theta d\phi^2),\label{metric1}
\end{equation}
 where%
\begin{equation}
 f(r)=\frac{4 Q^2 \, _2F_1\left(\frac{1}{4},\frac{1}{2};\frac{5}{4};-\frac{Q^2}{b^2 r^4}\right)}{3 r^2}+\frac{ 2 b^2 r^2}{3}
 \left(1-\sqrt{\frac{Q^2}{b^2 r^4}+1}\right)+\frac{r^2}{l^2}-\frac{2 M}{r}+1,\label{metric}
\end{equation}%
in which $_2F_1$
 is the hypergeometric function. From Eq.(\ref{metric}), we know that in the limit $b \rightarrow \infty$, $Q\neq0$,  the  solution reduces to the the Reissner-Nordstr\"{o}m-AdS  black hole, and in the  limit $Q\rightarrow 0$, it reduces to the Schwarzschild  AdS  black hole.

The ADM mass of the black hole, defined by $f(r_+) = 0$, is given by
\begin{equation}
M=\frac{4 l^2 Q^2 \, _2F_1\left(\frac{1}{4},\frac{1}{2};\frac{5}{4};-\frac{Q^2}{b^2 r_+^4}\right)-2 b^2 l^2 r_+^4 \sqrt{\frac{Q^2}{b^2 r_+^4}+1}+2 b^2 l^2 r_+^4+3 l^2 r_+^2+3 r_+^4}{6 l^2 r_+},\label{mass}
\end{equation}
in which $r_+$ is the event horizon of the black hole.
The Hawking temperature  defined by $T=\frac{f^{\prime}(r)}{4\pi}\mid_{r_+}$ can be written as
\begin{equation}
T=\frac{l^2 \left(1-2 b^2 r_+^2 \left(\sqrt{\frac{Q^2}{b^2 r_+^4}+1}-1\right)\right)+3 r_+^2}{4 \pi  l^2 r_+},\label{temperature}
 \end{equation}
where we have used Eq.(\ref{mass}).
In addition, according to the  Bekenstein-Hawking  entropy area relation, we can get the black hole entropy
\begin{equation}
S=\pi r_+^2.\label{entropy}
 \end{equation}
Inserting Eq.(\ref{entropy}) into Eq.(\ref{temperature}), we can get the entropy temperature relation, namely
\begin{equation}
T=\frac{-2 b^2 l^2 S \sqrt{\frac{\pi ^2 Q^2}{b^2 S^2}+1}+2 b^2 l^2 S+\pi  l^2+3 S}{4 \pi ^{3/2} l^2 \sqrt{S}}. \label{tentropy}
 \end{equation}
It is obvious that besides the charge $Q$, the  Born-Infeld parameter  $b$ also affects the phase structure of this space time in the $T-S$ plane.

\subsection{Phase transition of thermal entropy}
\begin{figure}
\centering
\subfigure[$Q=0.15$]{
\includegraphics[scale=0.75]{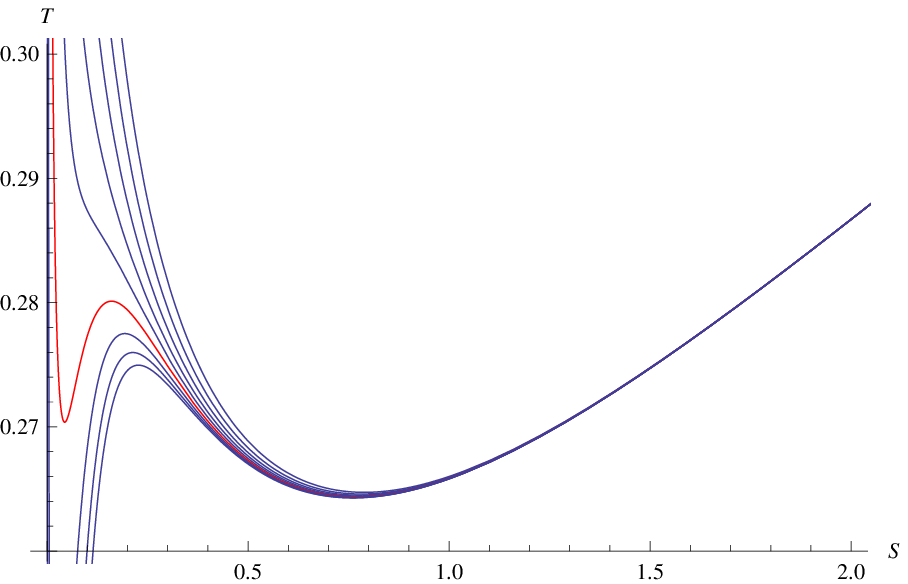}
}
\subfigure[$Q=0.2$]{
\includegraphics[scale=0.75]{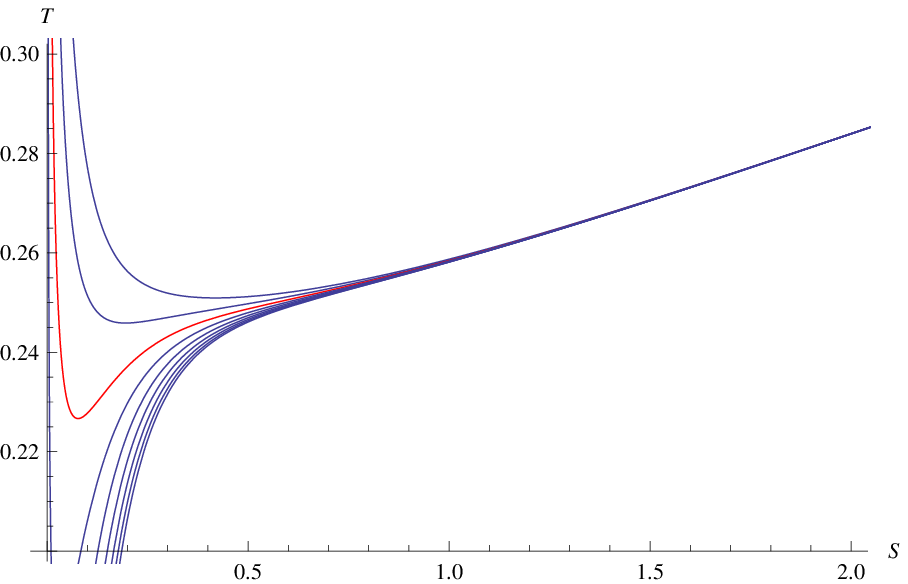}
}
\caption{\small Relations between entropy and temperature  for different $b$   at a fixed $Q$. The curves from top to down correspond to the case $b$ varies from 1.5 to 3.5 with step 0.25.} \label{fig1}
\end{figure}

\begin{figure}
\centering
\subfigure[$b=2$]{
\includegraphics[scale=0.75]{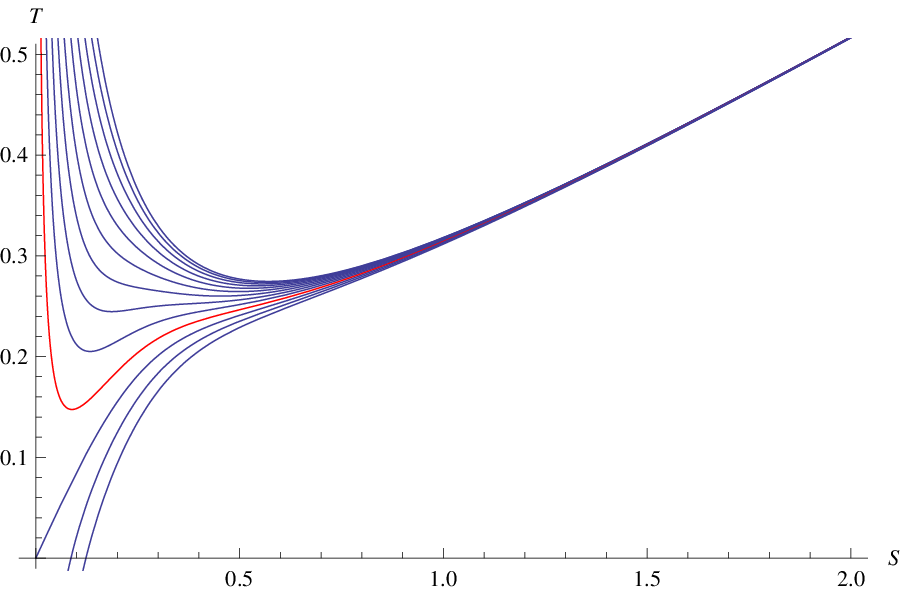}
}
\subfigure[$b=4$]{
\includegraphics[scale=0.75]{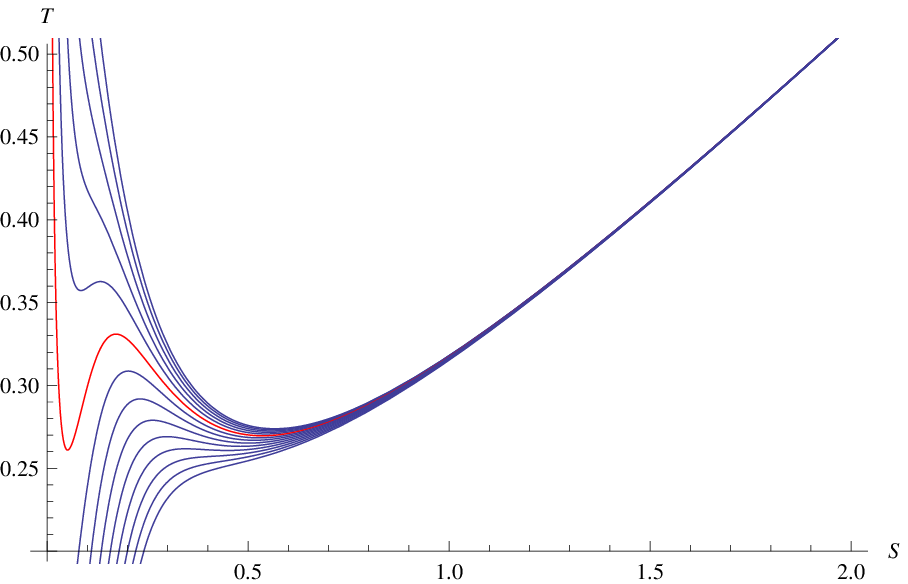}
}
\caption{\small Relations between entropy and temperature  for different $Q$   at a fixed $b$. The curves from top to down in (a) correspond to that $Q$ varies from 0.05 to 0.3 with step 0.02, and  in (b) they correspond to the case $Q$ varies from 0.065 to 0.195 with step 0.01.} \label{fig2}
\end{figure}

Based on Eq.(\ref{tentropy}), we will discuss the phase structure of the Born-Infeld-anti-de Sitter  black hole. We are
interested in
how  $b$  or  $Q$ affects it.
 For a fixed charge  $Q$,  the effect of $b$ on the phase structure is shown in Figure   \ref{fig1}. It is obvious that for the small $b$,  the phase structure resembles as that of the  Schwarzschild  AdS  black hole, and for large  $b$, it resembles as that of the Reissner-Nordstr\"{o}m-AdS  black hole as expected. From (b) of Figure \ref{fig1}, we can see that as  $b$ increases, the minimum temperature of the space time decreases, and  the entropy, also the event horizon, becomes smaller. That is, the  Born-Infeld parameter $b$ promotes the formation of an AdS space time.
More interestingly, for the smaller charge, we find there is a novel phase transition, which is labeled by the red solid line in (a) of Figure  \ref{fig1}. From this curve, we know that there are two unstable regions. Namely the black hole will undergo the following transition: unstable-stable-unstable-stable.  Compared with that of the  Reissner-Nordstr\"{o}m-AdS  black hole, a new branch emerges at the onset of the phase transition.
The influence  of the charge on the phase structure  for a fixed  Born-Infeld parameter is plotted in  Figure  \ref{fig2}.
Apparently,
 the phase structure  is similar to that of the  Schwarzschild  AdS  black hole  for the small charges, and Reissner-Nordstr\"{o}m-AdS  black hole for large  charges.
 From (a) of  Figure  \ref{fig2}, we observe that the larger the charge is, the lower the minimum temperature will be. In other words,  the charge will promote the formation of an AdS space time, which has the same effect as that of the Born-Infeld parameter on the phase structure of the black hole.
\begin{figure}
\centering
\subfigure[$b=5$]{
\includegraphics[scale=0.75]{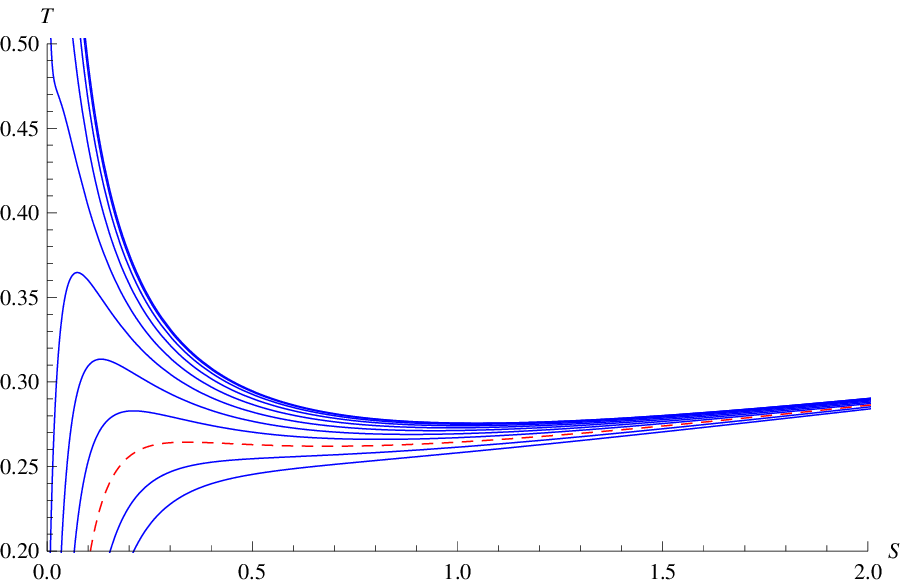}
}
\subfigure[$b=5$]{
\includegraphics[scale=0.75]{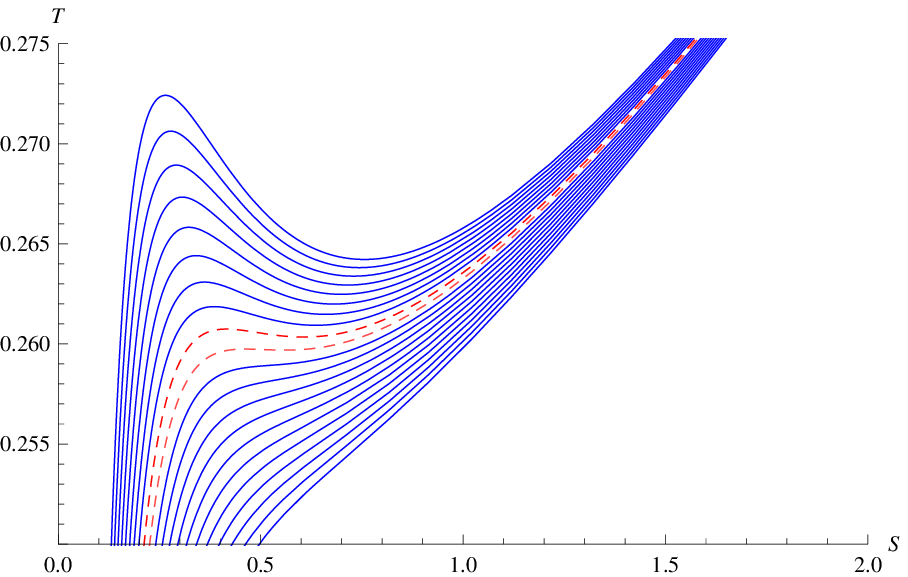}
}
\caption{\small Relations between entropy and temperature   for different $Q$   at a fixed $b$. The curves from top to down in (a) correspond to that $Q$ varies from 0.02 to 0.2 with step 0.02, and  in (b) they correspond to that $Q$ varies from 0.015 to 0.19 with step 0.002. } \label{fig3}
\end{figure}
For the case $b=4$ in (b) of Figure  \ref{fig2}, we  see that  for some intermediate values of charge $Q$, there is also a similar phase structure as that in (b) in Figure  \ref{fig1}, which is labeled by the red solid line in (b) of Figure  \ref{fig2}.  Note that however, for the large $b$, the novel phase structure disappears, which is shown in Figure  \ref{fig3}. For in this case, the  Born-Infeld parameter is large enough, so that the phase structure resembles completely as that of the Reissner-Nordstr\"{o}m-AdS  black hole.

Next, we  will study detailedly  the phase structure of the  Born-Infeld AdS black hole  for the case $b=4$ and  $b=5$ respectively. To finish it, we should first find the critical charge for a fixed Born-Infeld parameter by the condition \begin{equation}
(\frac{\partial T}{\partial S})_Q=(\frac{\partial^2 T}{\partial S^2})_Q=0. \label{heat}
 \end{equation}
 However, from  Eq.(\ref{temperature}), we find it is hard to get the analytical result directly. Taking the case  $b=5$ as an example,  we will show how to get it numerically next. We first plot a series of curves for different charges in the $T-S$ plane, which is shown in (a) of Figure  \ref{fig3}, and read off the charge which satisfies likely the condition  $(\frac{\partial T}{\partial S})_Q=0$. With that rough value, we plot a bunch of curves in the $T-S$ plane once with smaller step so that we can get the most likely value of charge.
  From (b) of Figure  \ref{fig3}, we find the critical charge should be $0.168<Q_c<0.17$, which are labeled by the red dashed lines in (b) of Figure  \ref{fig3}. Lastly, we adjust the value of $Q_c$ by hand to  find the only solution $S_c$ which satisfies   Eq.(\ref{heat}), which produces $Q_c=0.168678344129$, $S_c=0.510691$.
Substituting these critical values into Eq.(\ref{temperature}), we further get the critical temperature   $T_c=0.259444$.
 \begin{figure}
\centering
\subfigure[$b=5$]{
\includegraphics[scale=0.75]{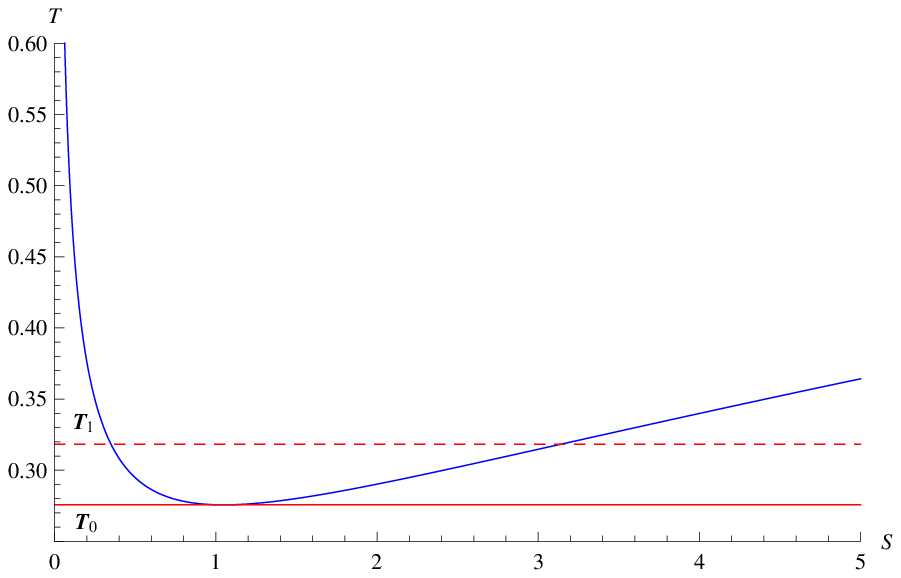}
}
\subfigure[$b=5$]{
\includegraphics[scale=0.75]{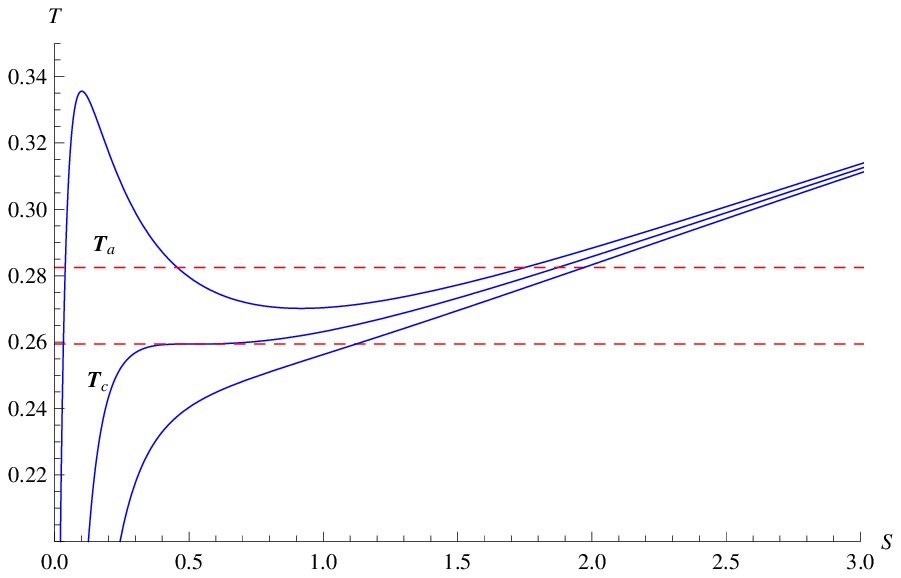}
}
\caption{\small Relations between entropy and temperature   for different $Q$  at $b=5$. The red solid line and dashed line in (a) correspond to the minimum temperature and Hawking-Page phase transition temperature. The red dashed lines from top to down in (b) correspond to the first order phase transition temperature and second order phase transition temperature.  } \label{fig4}
\end{figure}

\begin{figure}
\centering
\subfigure[$Q=0, b=5$]{
\includegraphics[scale=0.5]{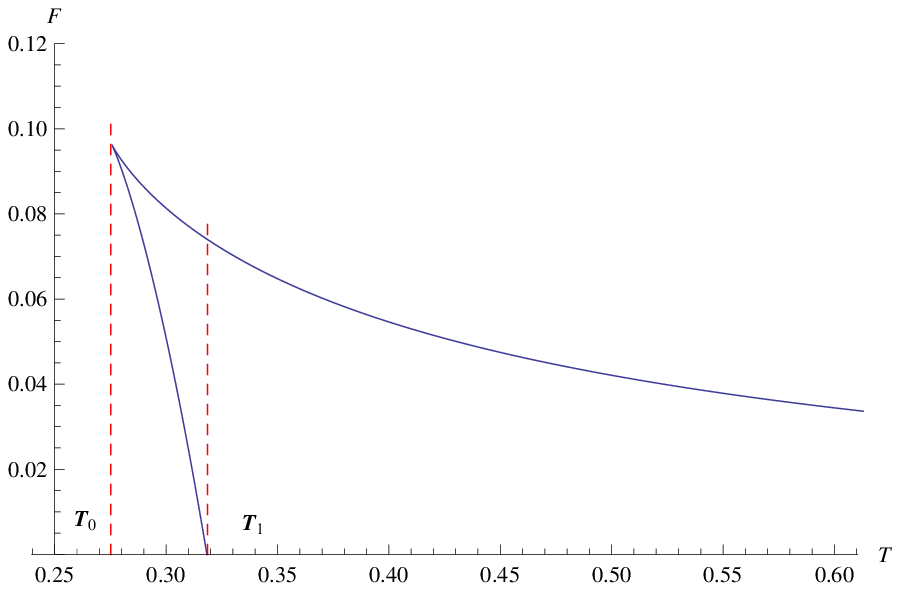}
}
\subfigure[$Q=1.1, b=5$]{
\includegraphics[scale=0.5]{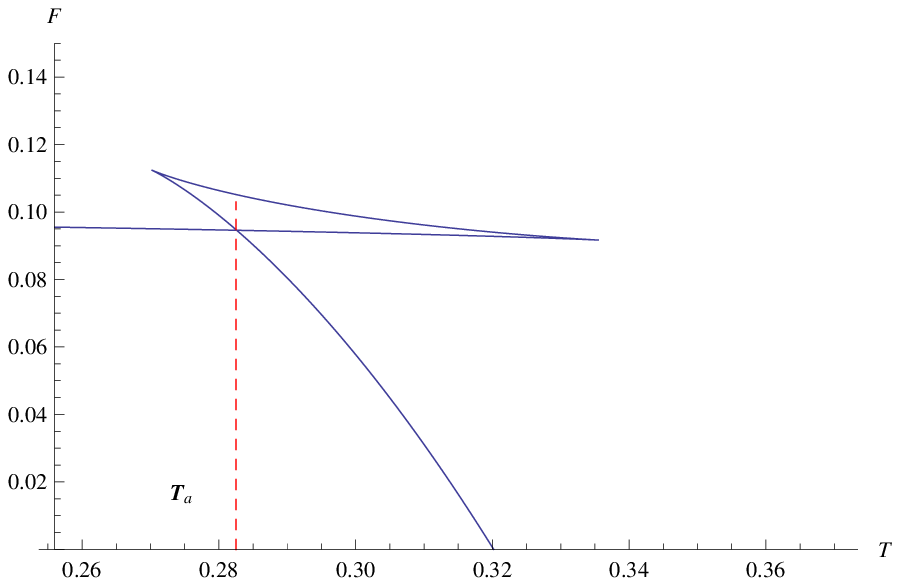}
}
\subfigure[$Q=0.168678344, b=5$]{
\includegraphics[scale=0.5]{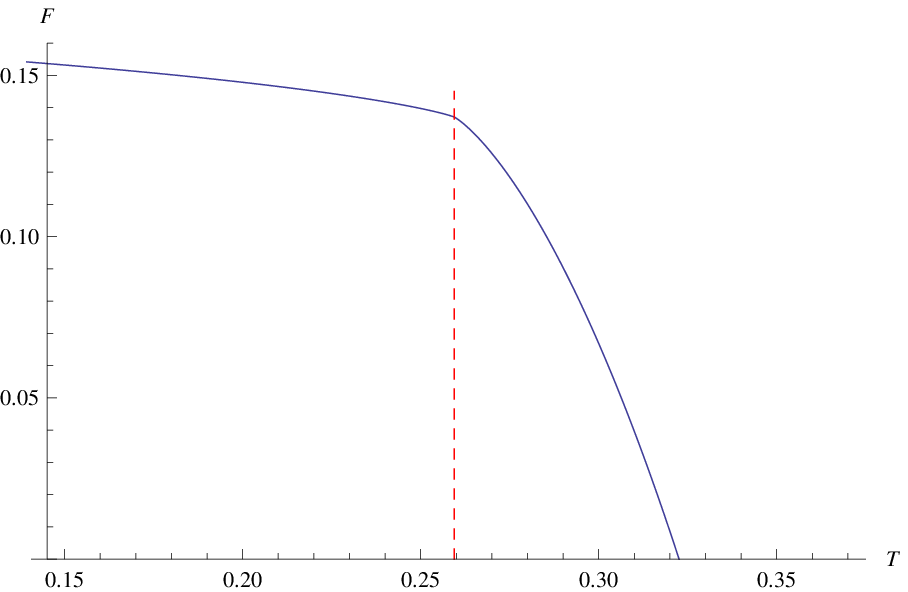}
}
\caption{\small Relations between the free energy and temperature for different charges at  $b=5$.  } \label{fig5}
\end{figure}

Having obtained the critical charge, we can plot the isocharge curves for the case $b=5$ in the $T-S$ plane, which is shown in  Figure  \ref{fig4}. We observe that the black hole undergoes a Hawking-Page phase transition and a Van der Waals-like phase transition.  Specifically speaking, for the case  $Q=0$, there
is a minimum temperature $T_0=\frac{\sqrt{3}}{2 \pi}$  \cite{Banerjee}, which is indicated by the red dashed line in (a) of Figure  \ref{fig4}.  When the temperature is higher
than $T_0$, there are two additional black hole branches. The small branch is unstable while the
large branch is stable.  The Hawking-Page phase transition occurs at the temperature given
by $T_1=\frac{1}{\pi}$  \cite{Banerjee}, which is indicated by the red dotted line. The  Hawking-Page phase transition  also can be observed in the $F-T$ plane, in which $F$ is the Helmholtz free energy defined by\footnote{Note that here we have implicitly chosen the pure AdS as the reference
spacetime because the free energy vanishes for pure AdS by this formula.}
\begin{equation}
F=\frac{8 l^2 Q^2 \, _2F_1\left(\frac{1}{4},\frac{1}{2};\frac{5}{4};-\frac{Q^2}{b^2  r_+^4}\right)+l^2  r_+^2 \left(2 b^2 r_+^2 \left(\sqrt{\frac{Q^2}{b^2  r_+^4}+1}-1\right)+3 \right)-3  r_+^4}{12 l^2  r_+}.
 \end{equation}
From (a) of Figure  \ref{fig5}, we know that there is a minimum temperature $T_0$, and above this temperature,  there are two branches.  The lower branch is stable always. The  Hawking-Page phase transition occurs at  $T_1$ for in this case the free energy vanishes.

For the case $Q\neq 0$, the phase structure is similar to that of the Van der Waals phase
transition, which is shown in (b) of Figure  \ref{fig4}. The solid blue lines from top to down correspond to the isocharges for the case $Q=0.11,  0.168678344129, 0.21$.
We can see that black holes endowed with different charges have different phase structures.
For the small charge, there is an unstable black hole interpolating between
the stable small hole and stable large hole.
 The small stable hole will jump to the large stable
hole at the critical  temperature $T_a$.  As the charge increases to the critical charge, the small hole and the large
hole merge into one and squeeze out the unstable phase so that an inflection point emerges. The divergence of the heat capacity in this case implies that the phase transition is second order.  As the charge exceeds the critical charge, we simply have one stable black hole
at each temperature.
The Van der Waals-like phase transition can also be observed from the $F-T$ relation. From  (b)
of Figure  \ref{fig5}, we observe  that there is a swallowtail structure, which corresponds to the unstable phase in the top curve in (b) Figure  \ref{fig4}.
The critical temperature $T_a=0.2825$ is apparently the value of  the horizontal coordinate of the junction between the small black hole and the large
black hole. As the temperature is lower than the critical temperature $T_a$, the free energy of
the small black hole is lowest, so the small hole is stable. As the temperature is higher than
$T_a$, the free energy of the large black hole is lowest, so that the large hole dominates thereafter.
The non-smoothness of the junction indicates that the phase transition is first order.
From (c) of Figure  \ref{fig5}, we know that there is an inflection point, which corresponds to the inflection point in the middle curve in  (b) of Figure  \ref{fig4}. The  horizontal coordinate of the inflection point  corresponds to the second order phase transition temperature $T_c$.

 \begin{figure}
\centering
\subfigure[$b=4$]{
\includegraphics[scale=0.75]{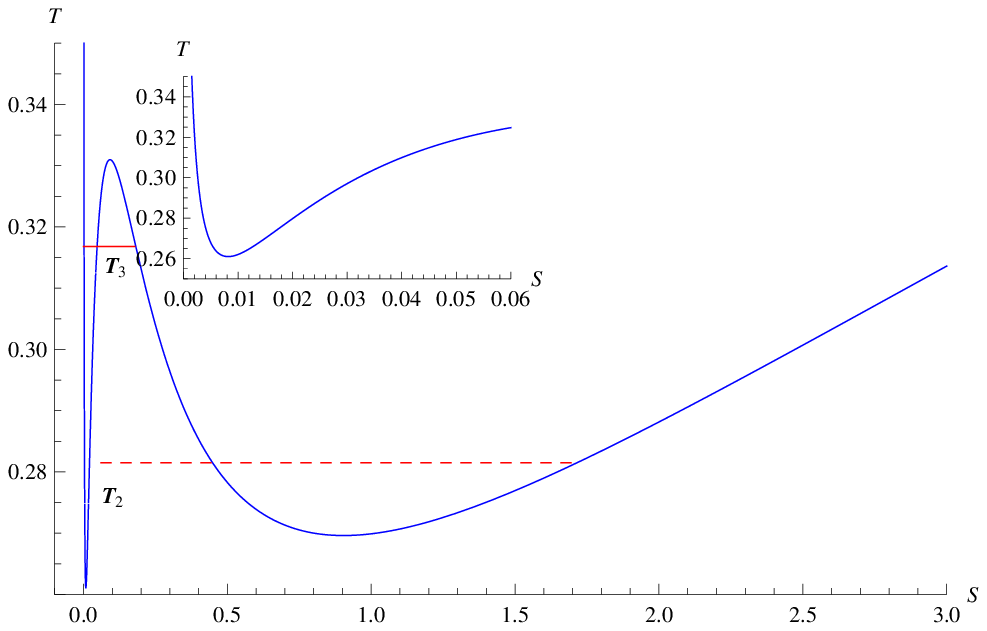}
}
\subfigure[$b=4$]{
\includegraphics[scale=0.75]{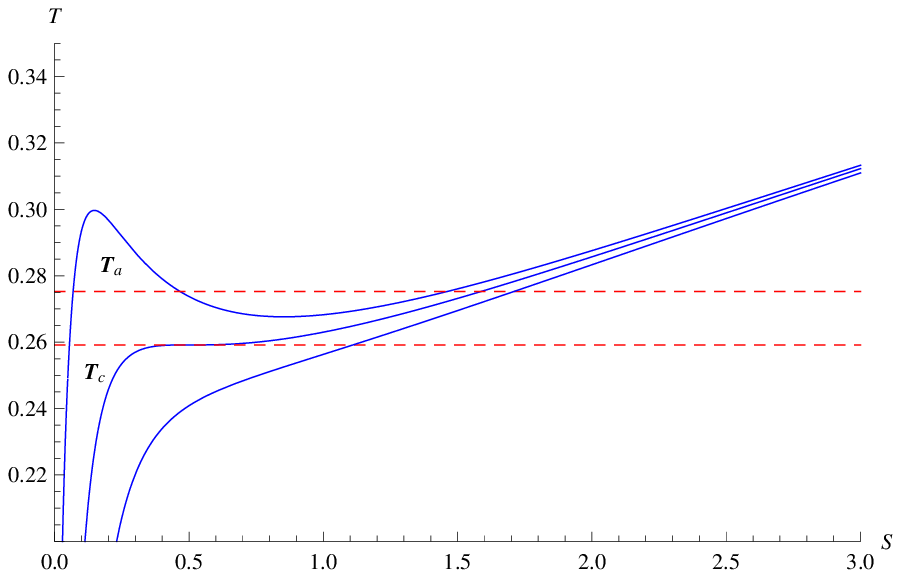}
}
\caption{\small Relations between entropy and temperature   for different $Q$   at $b=4$. The red  solid line and dashed  line in (a) correspond to pseudo phase transition temperature and first order  phase transition temperature. The red dashed lines from top to down in (b) correspond to the first order phase transition temperature and second order phase transition temperature.  } \label{fig6}
\end{figure}

\begin{figure}
\centering
\subfigure[$Q=0.115,b=4$]{
\includegraphics[scale=0.5]{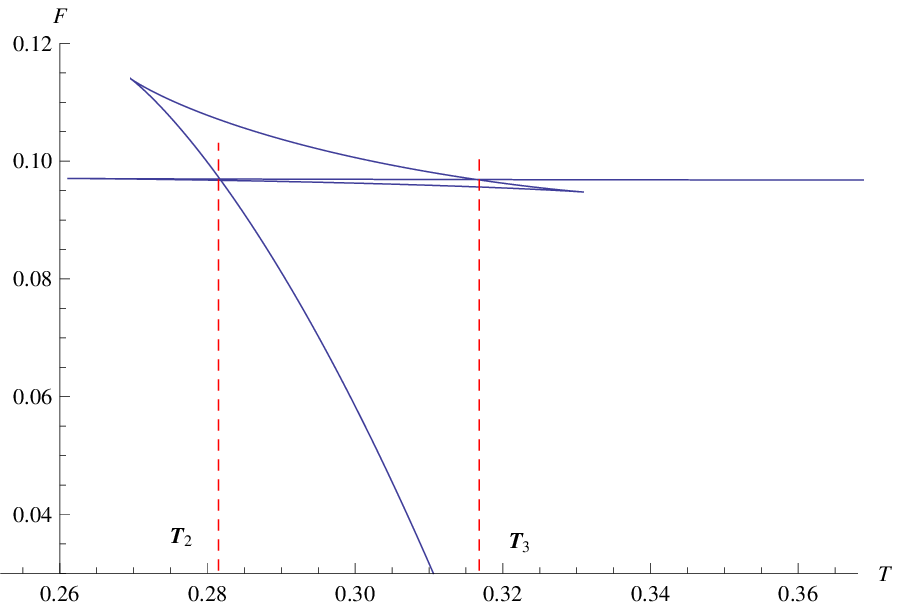}
}
\subfigure[$Q=0.13,b=4$]{
\includegraphics[scale=0.5]{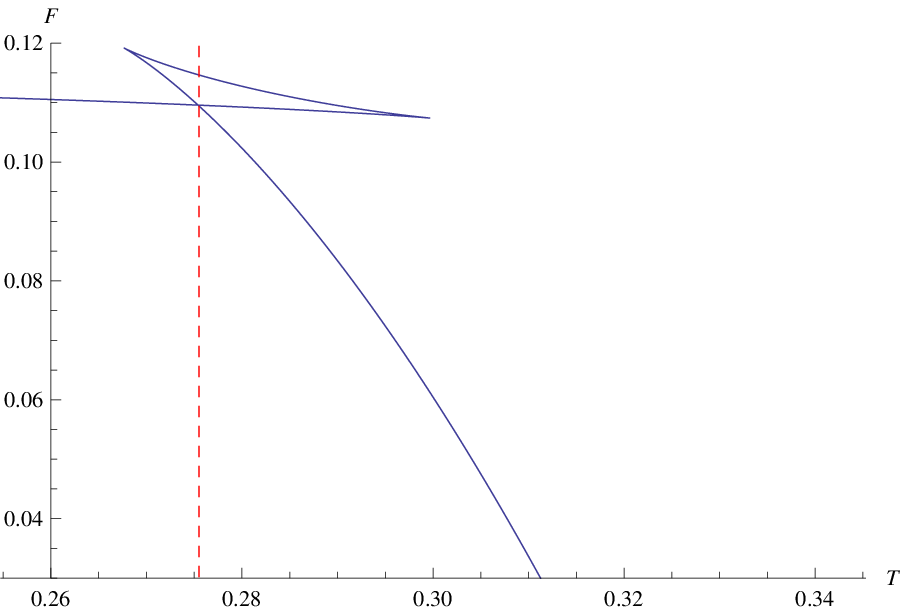}
}
\subfigure[$Q=0.169874523952,b=4$]{
\includegraphics[scale=0.5]{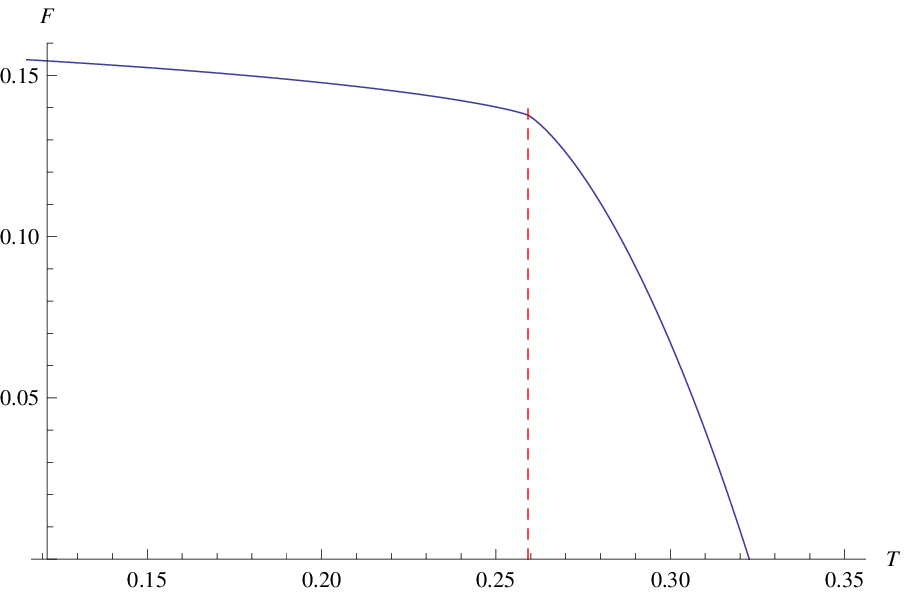}
}
\caption{\small Relations between the free energy and temperature for different charges at  $b=4$. } \label{fig7}
\end{figure}

Similarly, we also can  study the phase structure for the case $b=4$ in the $T-S$ plane.
To finish it, we should first find the critical charge. Adopting the same strategy as that of the case  $b=5$, we find $Q_c = 0.16987452395$,  $S_c = 0.502683752119928$,  $T_c =0.259166$. For the case $Q=0$, the phase structure is the same as that in  (a) of Figure  \ref{fig4}. We will not repeat it here. For the case $Q\neq 0$, we find besides the  Van der Waals-like phase
transition, there is a novel phase transition for $Q=0.115$ as observed in Figure  \ref{fig1} and Figure  \ref{fig2}, which is plotted in (a) of Figure  \ref{fig6}. That is, at the beginning of the phase transition, an  infinitesimally small black hole  branch emerges. It is unstable and its life  is very short so that it becomes to a small stable black hole quickly. In addition, we  find there are  two unstable region, and correspondingly in the $F-T$ plane, we observe two swallowtail structures, which is shown in (a) of  Figure  \ref{fig7}. The horizontal coordinate of the junction of the  swallowtail corresponds to the phase transition temperature, thus besides the first order phase transition temperature between the small black hole and large black hole, labeled by $T_2$, there is a new phase transition temperature, labeled by $T_3$.
 It should be stressed that the  new phase transition can not take place actually, for above the critical temperature $T_2$, the free energy of the large black hole is lowest so that the space time is dominated by the large black hole, which can be seen in (a) of  Figure  \ref{fig7}. Thus, we call thus phase transition pseudo phase transition.
We  observe that there is also a  Van der Waals-like phase
transition, which is plotted in (b) of Figure  \ref{fig6}.
 The solid blue lines from top to down correspond to the isocharges for the case $Q=0.13, 0.16987452395, 0.21$.
The first order phase transition temperature $ T_a$ can be read off from the horizontal coordinate of the junction of  the   swallowtail structure in (b) of  Figure  \ref{fig7}, and the second order phase transition temperature $ T_c$  can be read off from the horizontal coordinate of the inflection point in  (c) of  Figure  \ref{fig7}.

As done in \cite{Nguyen},  we will also check numerically whether Maxwell's equal area law holds for the first order phase transition and pseudo phase transition, which states
\begin{equation}
A_L\equiv\int_{S_{min}}^{S_{max}}T(S,Q)dS=T_x(S_{max}-S_{min})\equiv A_R \label{euqalarea},
 \end{equation}
in which $T(S,Q)$ is defined in  Eq.(\ref{tentropy}), $T_x$ is the phase transition temperature, and  $S_{min}$,  $S_{max}$ are the smallest and largest values of unstable region which satisfies the equation $T(S,Q)=T_x$. Usually for this equation, there are three roots $S_1,S_2,S_3$, so   $S_{min}=S_1$,  $S_{max}=S_3$, $T_x=T_a$.
But for the case $b=4,Q=0.115$, there are four roots. One can see from (a) of Figure  \ref{fig6} that $S_{min}=S_1$,  $S_{max}=S_3$  for the pseudo phase transition with  phase transition  temperature $T_x=T_3$, and  $S_{min}=S_2$,  $S_{max}=S_4$  for the  first order phase transition with   phase transition  temperature $T_x=T_2$.
The  calculated results for the first order phase transition and pseudo phase transition are listed in Table \ref{tab1}.
\begin{table}
\begin{center}\begin{tabular}{l|c|c|c|c|c|c|c|c
}
 \hline
                             &$S_{min}$ &          $S_{max}$     &$A_L$ &    $A_R$                   \\ \hline
$b=5, Q=0.11,T_a=0.2825$   & 0.039159   &  1.7506571   & 0.4837  &  0.4835          \\ \hline
$b=4,Q=0.13,T_a=0.2757$    & 0.0685377 &  1.46304     & 0.3840  &  0.3840 \\ \hline
$b=4,Q=0.115,T_2=0.2817$   & 0.0210019  &  1.72464    & 0.480 &  0.480 \\ \hline
$b=4, Q=0.115,T_3=0.3166$   &0.00212107 &  0.182767    & 0.05717 &  0.05719      \\ \hline
\end{tabular}
\end{center}
\caption{Check of the equal area law in the $T-S$ plane.}\label{tab1}
\end{table}
From this table, we can see that $A_L$ equals  $A_R$ for different $b$ in our numeric accuracy. The equal area law  therefore holds.

For the second order phase transition, we know that the heat capacity is divergent near the critical point and the critical exponent is $-2/3$. As stated in \cite{Johnson}, near the critical point, there is always a linear relation
\begin{eqnarray}
\log\mid T-T_c\mid =3 \log\mid S- S_c\mid +constant  \label{c2},
 \end{eqnarray}
with 3  the slope.  It is not difficult to show that for the case $b=4,5$ in our gravity model, the temperature and entropy also satisfy this linear relation near the critical point. Next, we will take Eq.(\ref{c2}) as a reference to check whether there is a similar relation for the  second order phase transition  in  the two point correlation function-temperature plane  as well as entanglement entropy-temperature plane.

\section{ Phase transition in the framework of holography}
\label{Nonlocal_observables}

Having obtained the phase structure of thermal entropy of the  Born-Infeld AdS black hole in the $T-S$ plane, we will study the phase structure of two point  correlation function  and  entanglement entropy in the fled theory to see whether they have the similar phase structure and critical behavior.

\subsection{Phase structure  probed by two point correlation function}

 According to the AdS/CFT correspondence, in the large  $\Delta$ limit, the equal time two point correlation function
 can be written as \cite{Balasubramanian61}
 \begin{equation}
\langle {\cal{O}} (t_0,x_i) {\cal{O}}(t_0, x_j)\rangle  \approx
e^{-\Delta {L}} ,\label{llll}
\end{equation}
where $\Delta$
  is  the conformal dimension of scalar operator $\cal{O}$ in the  dual field theory,
$L$ is the  length of the bulk geodesic between the points $(t_0,
x_i)$ and $(t_0, x_j)$ on the AdS boundary.
In our gravity model,
  we can simply choose $(\phi=\frac{\pi}{2},\theta=0)$ and $(\phi=\frac{\pi}{2},\theta=\theta_0)$ as the two boundary points. Then with  $\theta$ to
    parameterize the trajectory, the proper length  is given by
\begin{eqnarray}
L=\int_0 ^{\theta_0}\mathcal{L}(r(\theta),\theta) d\theta,~~\mathcal{L}=\sqrt{\frac{\dot{r}^2}{f(r)}+r^2},
 \end{eqnarray}
in which $\dot{r}=dr/ d\theta$.
Imagining $\theta$ as time, and treating $\mathcal{L}$ as the Lagrangian, one can
get the equation of motion for $r(\theta)$ by making use of the Euler-Lagrange equation. Then with the following  boundary conditions
\begin{eqnarray}
\dot{r}(0)=0, r(0)= r_0,\label{bon}
\end{eqnarray}
we can get the numeric result of $r(\theta)$.  To explore whether the size of the boundary region affects the phase structure,  we  will choose  $\theta_0=0.2, 0.3$  as two examples. Note that for a fixed $\theta_0$, the geodesic length  is divergent, so it should be regularized by subtracting off the geodesic length in pure AdS with the same  boundary region, denoted by $L_0$\footnote{$L_0$ can be obtained easily for there is an analytical result for $r(\theta)$ in the pre AdS, namely $r_{AdS}(\theta)=l[(\frac{\cos \theta}{\cos \theta_0})^2-1]^{-1/2}$ \cite{Blanco, Casini}. }. To achieve this, we are required to set a UV cutoff for each case, which is chosen to be $ r(0.199)$ and $ r(0.299)$, respectively for our two examples.
The regularized  geodesic length is labeled as $\delta L\equiv L-L_0$. In addition, during the numerics, we will set the AdS radius $l$   to be  1.

\begin{figure}
\centering
\subfigure[$b=5, \theta_0=0.2$]{
\includegraphics[scale=0.75]{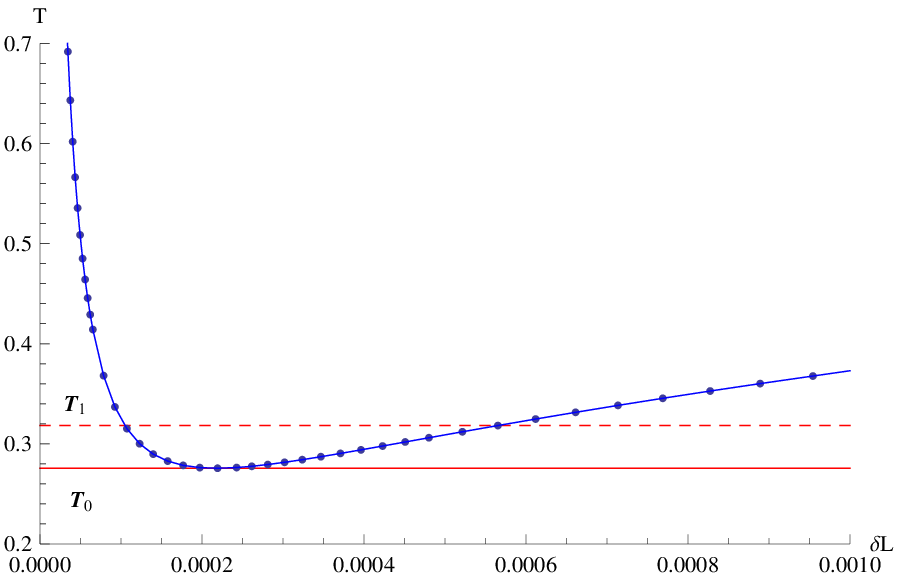}
}
\subfigure[$b=5, \theta_0=0.2$]{
\includegraphics[scale=0.75]{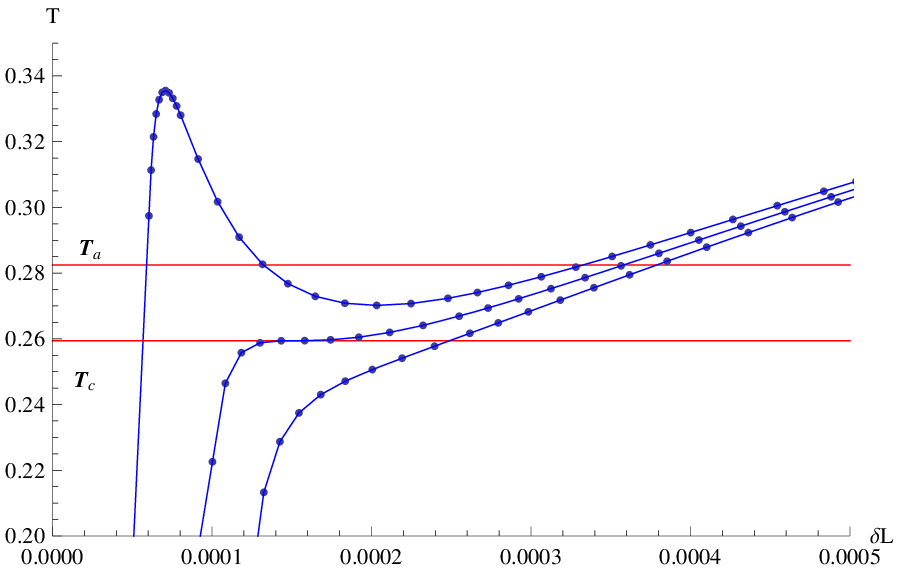}
}
\caption{\small  Isocharges in the $\delta L-T$ plane for the case $b=5, \theta_0=0.2$. The red solid line and dashed line in (a) correspond to the minimum temperature and Hawking-Page phase transition temperature. The red  solid lines from top to down in (b) correspond to the first order phase transition temperature and second order phase transition temperature.  } \label{fig8}
\end{figure}

\begin{figure}
\centering
\subfigure[$b=5, \theta_0=0.3$]{
\includegraphics[scale=0.75]{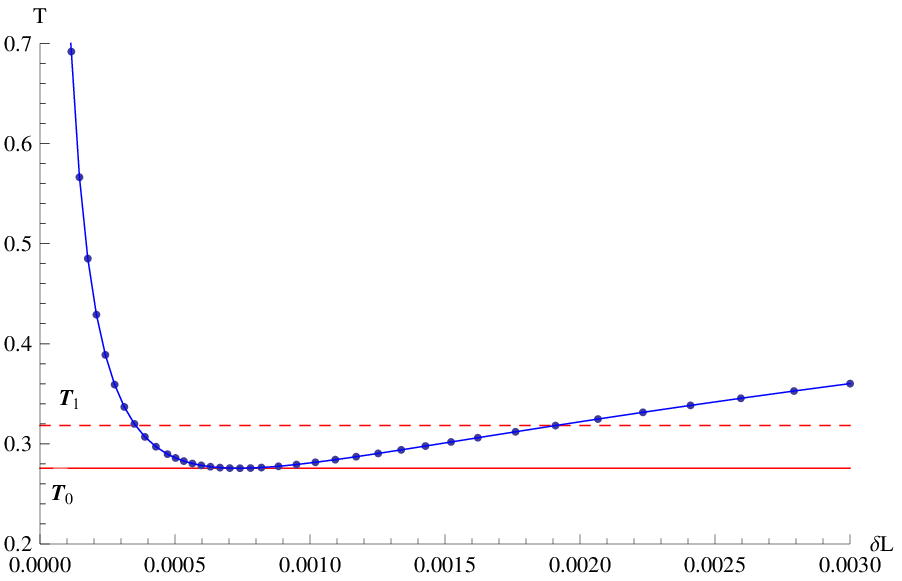}
}
\subfigure[$b=5, \theta_0=0.3$]{
\includegraphics[scale=0.75]{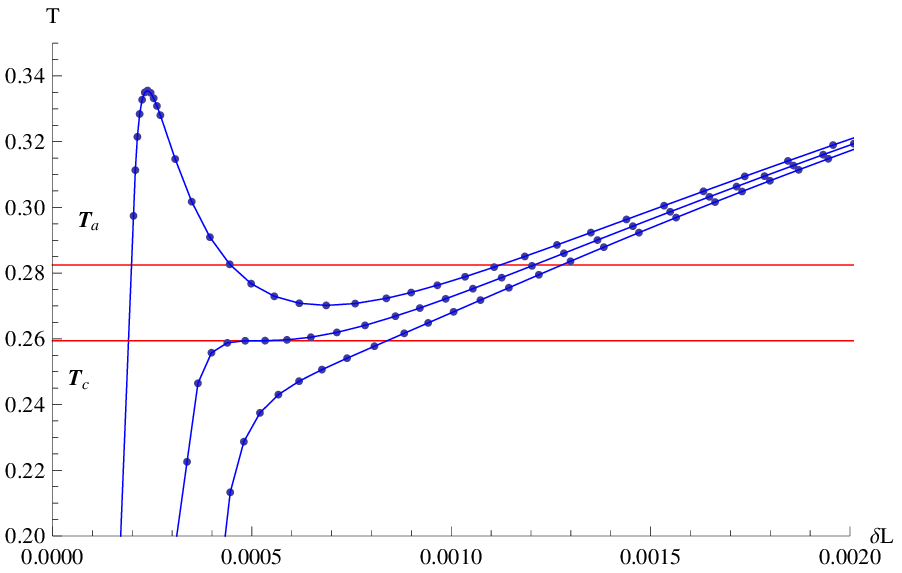}
}
\caption{\small  Isocharges in the $\delta L-T$ plane for the case $b=5, \theta_0=0.3$. The red solid line and dashed line in (a) correspond to the minimum temperature and Hawking-Page phase transition temperature. The red solid lines from top to down in (b) correspond to the first order phase transition temperature and second order phase transition temperature.  } \label{fig9}
\end{figure}

\begin{figure}
\centering
\subfigure[$b=4, \theta_0=0.2$]{
\includegraphics[scale=0.75]{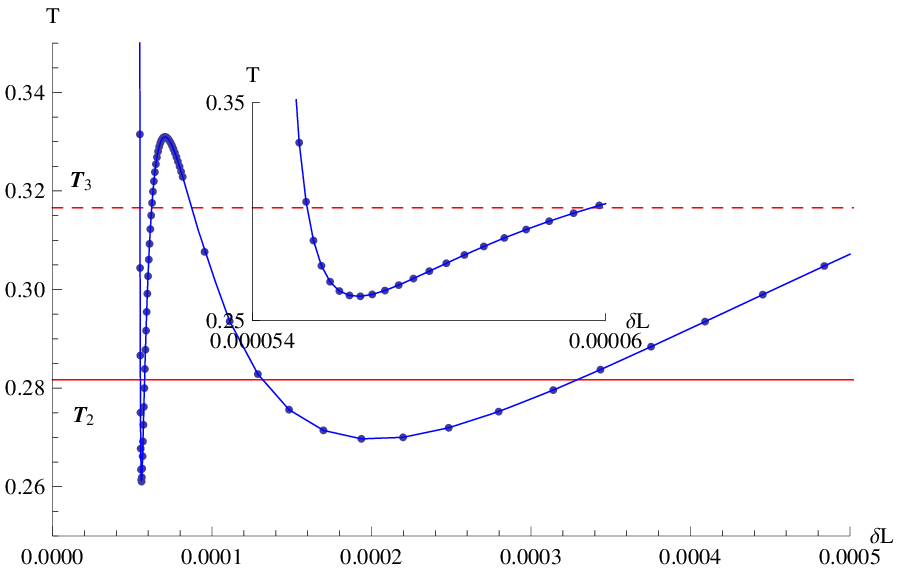}}
\subfigure[$b=4, \theta_0=0.2$]{
\includegraphics[scale=0.75]{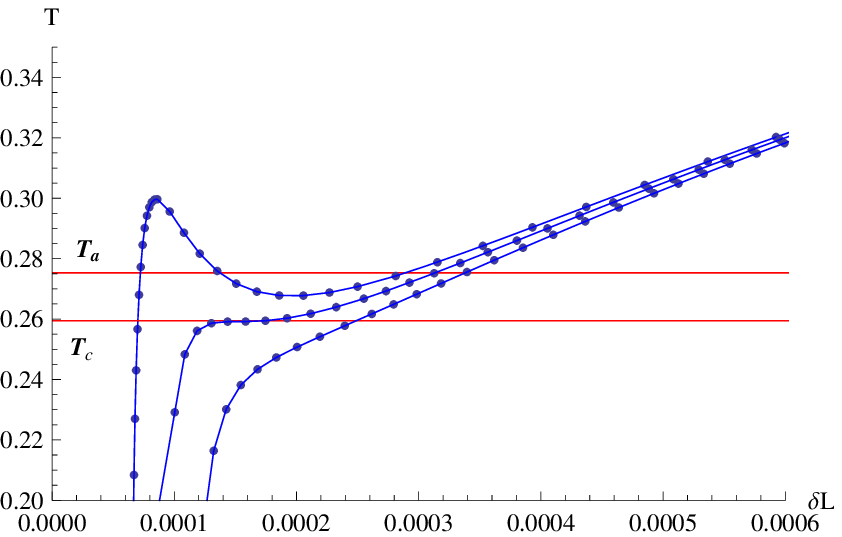}
}
\caption{\small  Isocharges in the $\delta L-T$ plane for the case $b=4, \theta_0=0.2$.  The red solid line and dashed line in (a) correspond to the minimum temperature and Hawking-Page phase transition temperature. The red solid  lines from above to down in (b) correspond to the first order phase transition temperature and second order phase transition temperature.  } \label{fig10}
\end{figure}

\begin{figure}
\centering
\subfigure[$b=4$]{
\includegraphics[scale=0.75]{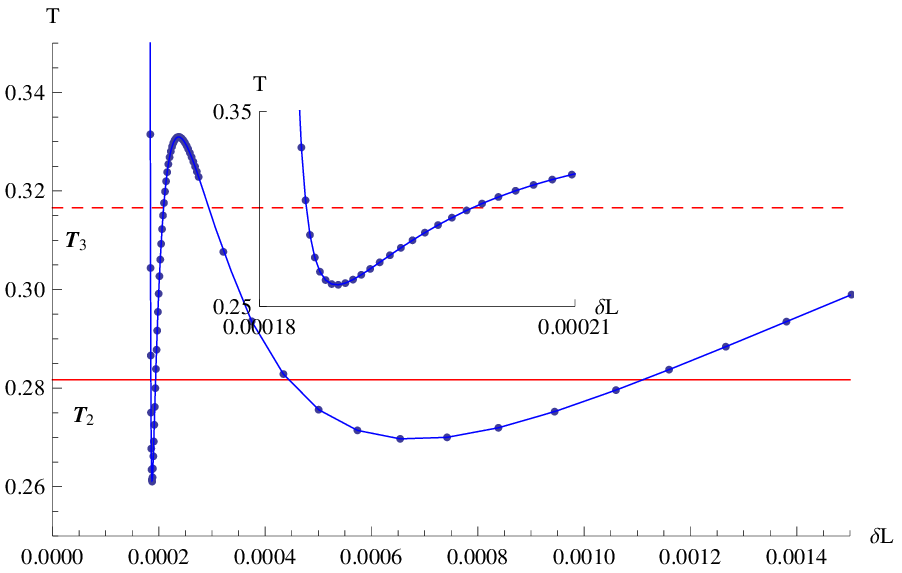}
}
\subfigure[$b=4$]{
\includegraphics[scale=0.75]{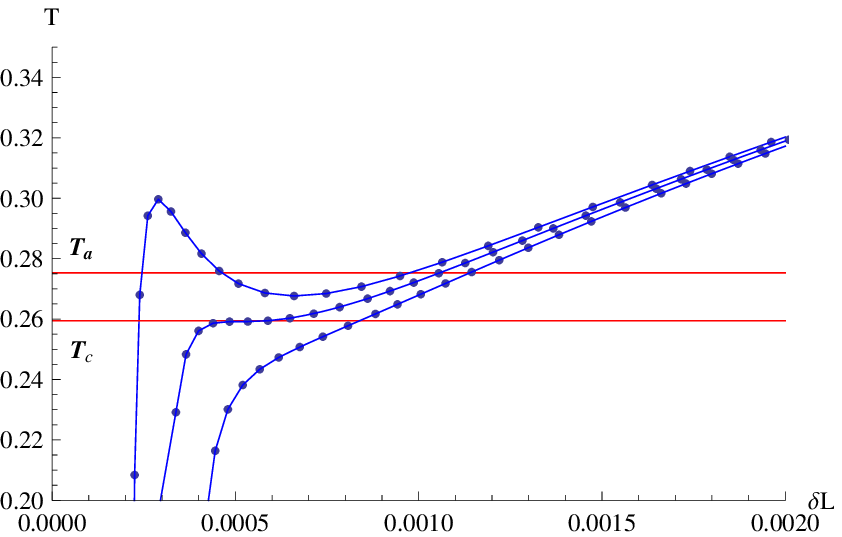}
}
\caption{\small Isocharges in the $\delta L-T$ plane for the case $b=4, \theta_0=0.3$.  The red solid line and dashed line in (a) correspond to the minimum temperature and Hawking-Page phase transition temperature. The red solid lines from top to down in (b) correspond to the first order phase transition temperature and second order phase transition temperature.  } \label{fig11}
\end{figure}

We plot the relation between  $T$ and $\delta L$ for the case $b=5$ in Figure \ref{fig8} and  Figure \ref{fig9} . The solid blue line in (a) corresponds to the isocharge for the case $Q=0$, and the dashed blue lines from top to down in (b) correspond  to that  $Q=0.11,  0.168678344129, 0.21$.
 From these figures, one can see that  $\delta L$ demonstrates a similar phase structure as that of the thermal entropy in Figure \ref{fig4}. Namely the two point correlation function can probe both the Hawking-Page phase transition and
 Van der Waals-like phase transition.
Precisely speaking,
 we find that  the minimum temperature $T_0$ as well as Hawking-Page phase transition temperature  $T_1$ in $(a)$,  the first order phase transition temperature  $T_{a}$, and second order phase transition temperature  $T_c$ in $(b)$ are exactly the same as those in $T-S$ plane. This conclusion will not be affected  as  $\theta_0$  varies in a reasonable region, which can be seen from Figure \ref{fig8} and  Figure \ref{fig9}.

With the two point correlation function, we also can probe the phase structure of the case $b=4$, which is shown in  Figure \ref{fig10} and  Figure \ref{fig11}.   The solid blue line in (a) corresponds to the isocharge for the case $Q=0.115$, and the solid blue lines from top to down in (b) correspond  to that $Q=0.13, 0.16987452395, 0.21$. As the case as  the thermal entropy in the $T-S$ plane, we find there ia also a novel phase structure besides the  Van der Waals-like phase transition  in the $\delta L-T$  plane. There are also two unstable region, correspondingly  two
phase transition temperature, labeled by  $T_2$,  $T_3$.
For the phase transition temperature mentioned above,
  it is easy to check $T_0$ by locating the position of local minimum. But in order to locate  $T_2$, $T_3$, and $T_{a}$ precisely, we are required to examine the equal area law for the first order phase transition as well as pseudo phase transition. And to locate $T_c$, we should  obtain the critical exponent  $-2/3$ for the second order phase transition, which are documented as follows.

  In the  $\delta L-T$ plane, we define the  equal area law  as
\begin{eqnarray}
A_L\equiv\int_{\delta L_{min}}^{\delta L_{max}} T(\delta L) d\delta L= T_{x} (\delta L_{max}-\delta L_{min})\equiv A_R,\label{arealaw}
 \end{eqnarray}
in which  $T(\delta L)$  is an Interpolating Function obtained from the numeric result, $T_{x} $ is the phase transition temperature, and $\delta L_{min}$,  $\delta L_{max}$ are the smallest and largest values of the unstable region which satisfy the equation $T(\delta L)=T_{x}$. Similar to the equal area law in the  $S-T$  plane, we
know that for $b=5,Q=0.11$ and $b=4,Q=0.13$, $T_{x}=T_{a}$, while for $b=4,Q=0.115$, $T_{x}=T_{2}$ for the first order phase transition and $T_{x}=T_{3}$ for the pseudo phase transition.
For different $b$ and $\theta_0$, the calculated results are listed in Table  \ref{tab2}.
\begin{table}
\begin{center}\begin{tabular}{l|c|c|c|c|c|c|c|c}
 \hline
     &$\delta L_{min}$ &          $\delta L_{max}$     &$A_L$ &    $A_R$                   \\ \hline
$b=5, Q=0.11,T_a=0.2825,\theta_0=0.2$   &0.00005901   &0.0003329  &0.0000777 & 0.0000774     \\ \hline
$b=5, Q=0.11,T_a=0.2825,\theta_0=0.3$   & 0.0001982  & 0.001124  & 0.000263  &  0.000261    \\ \hline
$b=4, Q=0.13,T_a=0.2757,\theta_0=0.2$  & 0.00007227 &  0.0002890    & 0.0000598  & 0.0000596  \\ \hline
$b=4, Q=0.13,T_a=0.2757,\theta_0=0.3$    &0.0002453 &  0.0009755   & 0.000201  & 0.000201 \\ \hline
$b=4, Q=0.115,T_2=0.2817,\theta_0=0.2$   &0.00005509 &  0.0003290    & 0.0000774  &0.0000772 \\ \hline
$b=4, Q=0.115,T_2=0.2817,\theta_0=0.3$   & 0.0001939 &  0.001110    & 0.000259  &  0.000258\\ \hline
$b=4, Q=0.115,T_3=0.3166,\theta_0=0.2$   & 0.00005485 &  0.00008727   &0.0000103 &  0.0000103   \\ \hline
$b=4, Q=0.115,T_3=0.3166,\theta_0=0.3$   & 0.0001842 &  0.00003476   &0.0000349 &  0.0000348    \\ \hline
\end{tabular}
\end{center}
\caption{Check of the equal area law in the $T-\delta L$ plane.}\label{tab2}
\end{table}
It is obvious that $A_L$ equals  $A_R$ for different $b$ and $\theta_0$ in a  reasonable accuracy. That is, in the $T-\delta L$ plane, the equal area law holds, and it is independent of the Born-Infeld parameter as well as the size of the boundary region.

\begin{figure}
\centering
\subfigure[$b=5, \theta_0=0.2$]
{\includegraphics[scale=0.75]{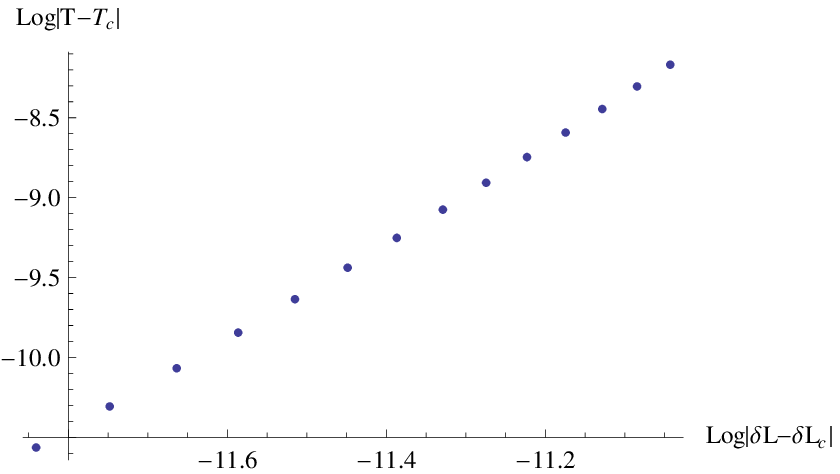}  }
\subfigure[$b=5, \theta_0=0.3$]
{\includegraphics[scale=0.75]{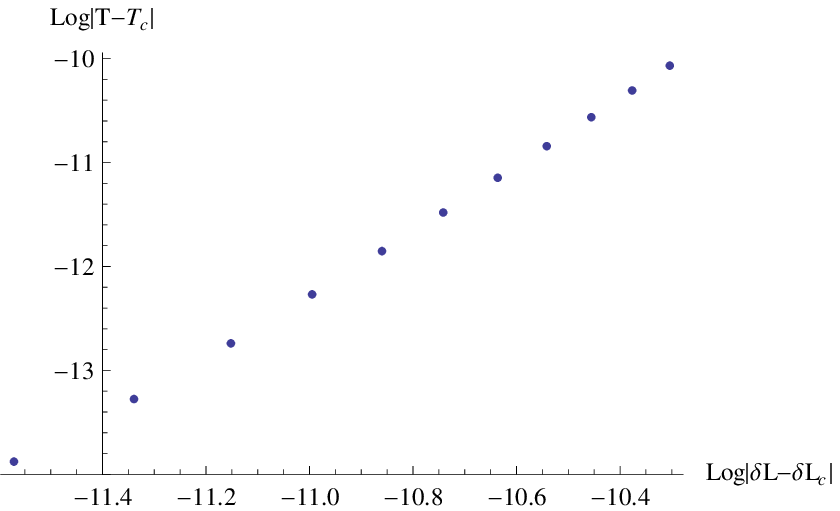}  }
\subfigure[$b=4, \theta_0=0.2$]{
\includegraphics[scale=0.75]{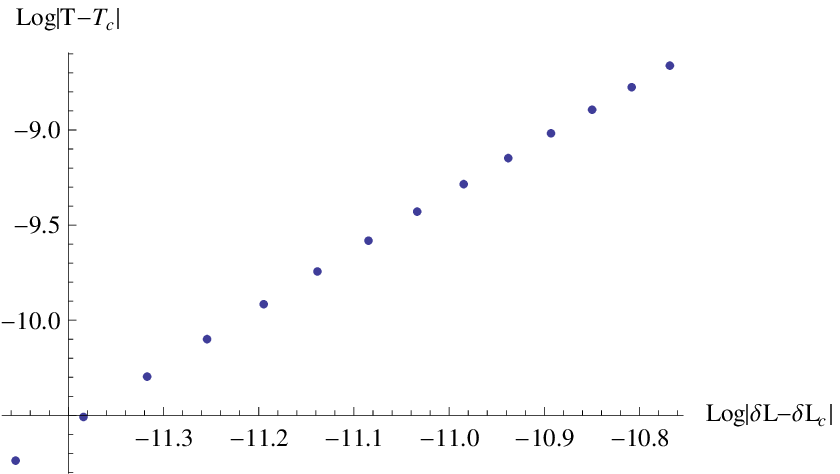}  }
\subfigure[$b=4,\theta_0=0.3$]
{\includegraphics[scale=0.75]{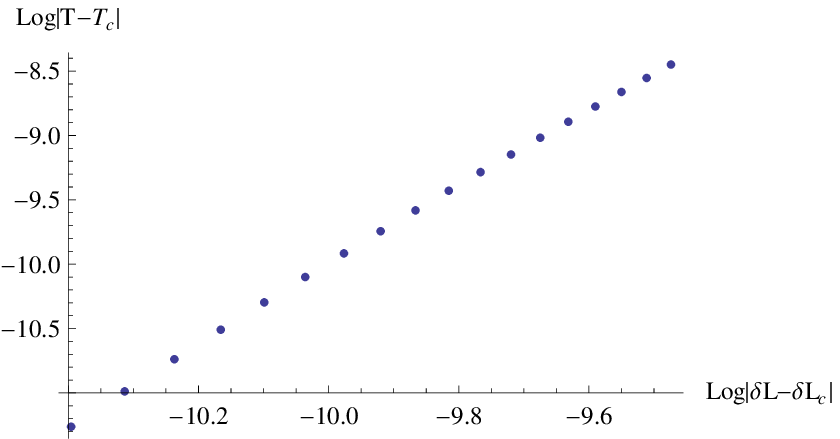}  }
\caption{\small  Relations between $\log\mid T-T_c\mid$ and  $\log\mid\delta L-\delta L_c\mid $ for different $b$ and $\theta_0$.} \label{fig12}
\end{figure}

In order to get the critical exponent for the second order phase transition in the $T-\delta L$  plane, we should first define an analogous heat capacity
\begin{eqnarray}
C=T\frac{\partial \delta L}{\partial T}. \label{cheat2}
 \end{eqnarray}
Provided a similar relation as showed  in Eq.(\ref{c2}) is satisfied, one can get the critical exponent immediately.
So next, we are interested in the logarithm of the quantities $T-T_c$, $\delta L-\delta L_c$, in which $T_c$ is the critical temperature mentioned previously, and
 $L_c$ is  obtained numerically by the equation $T(\delta L)=T_c$.
 We plot the relation between $ \log\mid T -T_c\mid$ and $\log\mid\delta L-\delta L_c\mid  $ for different $\theta_0$  and $b$ in Figure \ref{fig12}, where
 these
straight lines can be fitted as
\begin{equation}
\log\mid T-T_c\mid=\begin{cases}
25.0567 + 3.0113    \log\mid\delta L-\delta L_c\mid,&$for$~b=5, \theta_0=0.2,\\
21.207 + 3.0402   \log\mid\delta L-\delta L_c\mid,&  $for$ ~b=5, \theta_0=0.3,\\
23.8282+3.01531   \log\mid\delta L-\delta L_c\mid, & $for$~b=4, \theta_0=0.2,\\
20.3704 + 3.03817   \log\mid\delta L-\delta L_c\mid,& $for$ ~b=4,\theta_0=0.3.\\
\end{cases}
\end{equation}
It is obvious that the slope is about 3, which indicates that the critical exponent is $-2/3$ nearly for the analogous heat capacity, and the phase transition is also second order with the phase transition temperature $T_c$.

\subsection{Phase structure   probed by holographic entanglement entropy}

 According to the formula in \cite{Ryu,Ryu1}, holographic entanglement entropy can be given by the area
$A_{\Sigma}$ of a minimal surface $\Sigma$ anchored on the boundary entangling surface $\partial \Sigma$, namely
\begin{equation}
S=\frac{A_{\Sigma}(t)}{4 G}, \label{eee}
\end{equation}
where $G$ is the  Newton's constant.  We will take the region $\Sigma$ to be a spherical cap on the boundary delimited by $\theta\leq \theta _0$\footnote{To avoid contamination of the entanglement entropy by the thermal entropy,  we will choose a small entangling region as pointed out in  \cite{Johnson,Hubeny}.  }. Then based on the definition of area and  (\ref{metric1}),
(\ref{eee}) can be rewritten as
\begin{eqnarray}
S=\frac{ \pi}{2} \int_0 ^{\theta_0}r \sin\theta\sqrt{\frac{(r^{\prime})^2}{f(r)}+r^2},
 \end{eqnarray}
in which $r^{\prime}=dr/ d\theta$.
 Making use of the Euler-Lagrange equation, one can get the equation of motion of $r(\theta)$. With the boundary conditions in  Eq.(\ref{bon}), we can obtain the numeric result of  $r(\theta)$ immediately. Note that the entanglement entropy is divergent at the boundary, so it should be regularized by subtracting off the entanglement entropy in pure AdS with the same entangling surface and boundary values.
We label the regularized entanglement entropy as $\delta S$.
We choose the size of the boundary region as  $\theta_0=0.2$ and set the UV cutoff in the dual field theory  to be $r(0.199)$.

\begin{figure}
\centering
\subfigure[$b=5, \theta_0=0.2$]{
\includegraphics[scale=0.75]{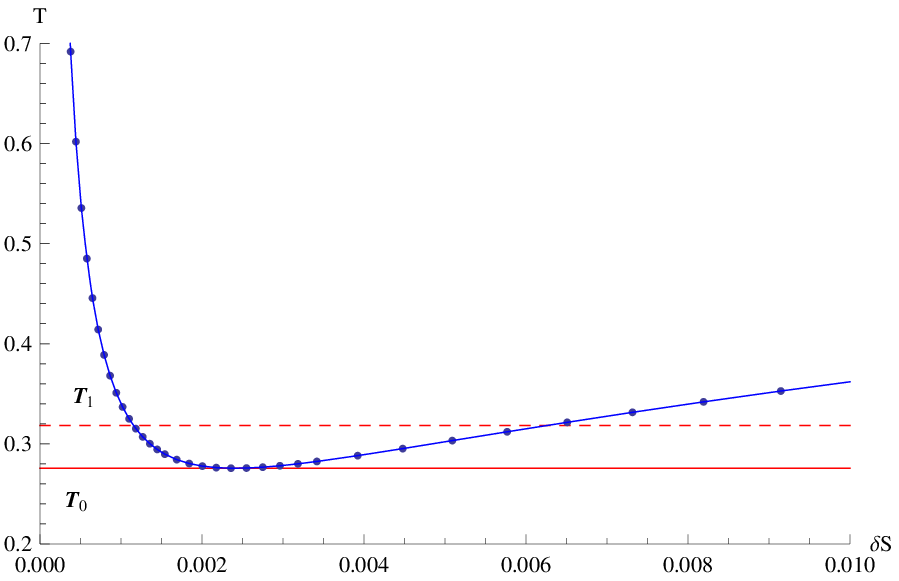}
}
\subfigure[$b=5, \theta_0=0.2$]{
\includegraphics[scale=0.75]{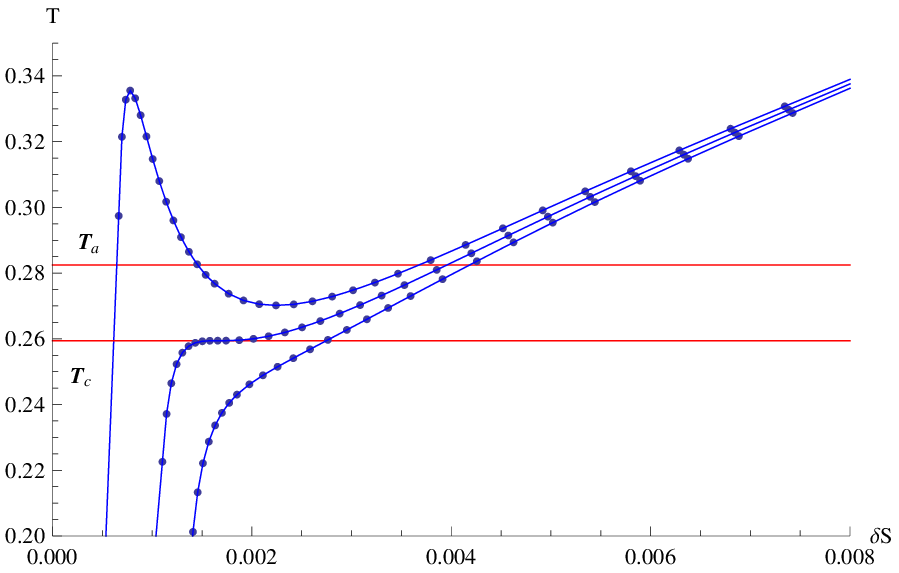}
}
\caption{\small  Isocharges in the $\delta S-T$ plane for the case $b=5, \theta_0=0.2$. The red solid line and dashed line in (a) correspond to the minimum temperature and Hawking-Page phase transition temperature. The red  solid lines from top to down in (b) correspond to the first order phase transition temperature and second order phase transition temperature.  } \label{fig13}
\end{figure}

\begin{figure}
\centering
\subfigure[$b=4, \theta_0=0.2$]{
\includegraphics[scale=0.75]{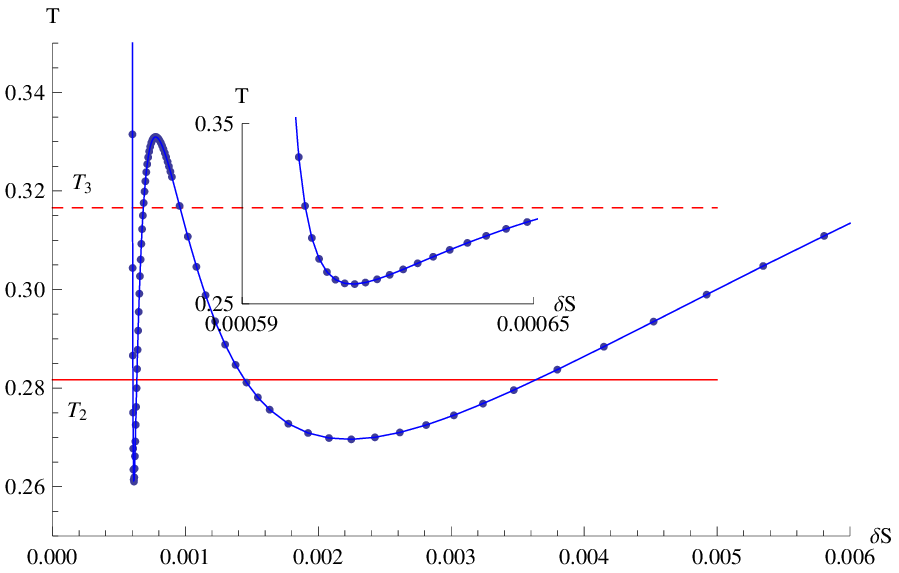}}
\subfigure[$b=4, \theta_0=0.2$]{
\includegraphics[scale=0.75]{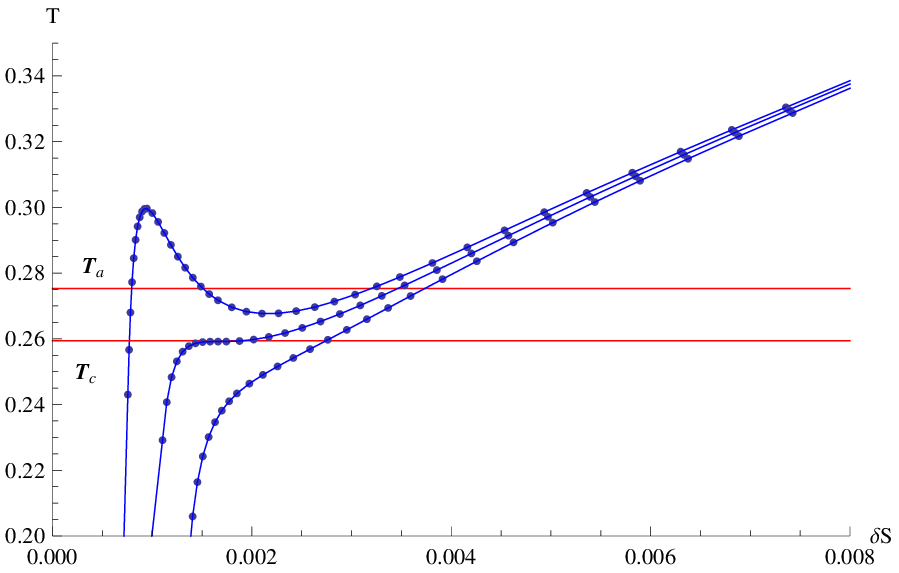}
}
\caption{\small  Isocharges in the $\delta S-T$ plane for the case $b=4, \theta_0=0.2$.  The red solid line and dashed line  in (a) correspond to the minimum temperature and Hawking-Page phase transition temperature. The red solid  lines from above to down in (b) correspond to the first order phase transition temperature and second order phase transition temperature.  } \label{fig14}
\end{figure}

The isocharges for the cases $b=5$ and  $b=4$  on the  entanglement entropy-temperature plane are plotted in  Figure \ref{fig13} and  Figure \ref{fig14}.  As the same as that of the two point correlation function, for  $b=5$, we are interested in that the black holes have charges $Q=0, 0.11,  0.168678344129, 0.21$, and for  $b=4$, we choose $Q=0.115, 0.13, 0.16987452395, 0.21$. From Figure \ref{fig13}, we know that the entanglement entropy exhibits a Hawking-Page phase transition,  and a Van der Waals-like phase transition as the charge increases form 0 to 0.21. The entanglement entropy also exhibits the novel phase structure as that of the thermal entropy as well as two point correlation function, which is shown in (a) of  Figure \ref{fig14}.

We will also employ  Maxwell's equal area law to locate the first order phase transition temperature, namely $T_a$ in (b) of  Figure \ref{fig13} and  Figure \ref{fig14}, $T_2$ in (a) of  Figure \ref{fig14}, and pseudo phase transition temperature $T_3$.
In the  $\delta S-T$ plane, we define the  equal area law as
\begin{eqnarray}
A_L\equiv\int_{\delta S_{min}}^{\delta S_{max}} T(\delta S) d\delta S= T_{x} (\delta S_{max}-\delta S_{min})\equiv A_R,\label{arealaw}
 \end{eqnarray}
in which  $T(\delta S)$  is an Interpolating Function obtained from the numeric result, and $\delta S_{min}$,  $\delta S_{max}$ are the smallest and largest roots of the unstable regions which satisfy the equation $T(\delta S)=T_{x}$. Surely the values of the phase transition temperature $ T_{x} $ depends on the choice of $Q$ and $b$.
For different $Q$ and $b$, the results of  $\delta S_{min}$,  $\delta S_{max}$  and $A_L$,  $A_R$ are listed in Table
(\ref{tab3}).
\begin{table}
\begin{center}\begin{tabular}{l|c|c|c|c|c|c|c|c}
 \hline
     &$\delta S_{min}$ &          $\delta S_{max}$     &$A_L$ &    $A_R$                   \\ \hline
$b=5, Q=0.11,T_a=0.2825,\theta_0=0.2$   &0.00064474  &0.0036781  &0.000860& 0.000857    \\ \hline
$b=4, Q=0.13,T_a=0.2757,\theta_0=0.2$  &0.00076078&  0.0032308& 0.000681  &0.000681  \\ \hline
$b=4, Q=0.115,T_2=0.2817,\theta_0=0.2$   &0.00060499 & 0.0036363   & 0.000857 &  0.000853 \\ \hline
$b=4, Q=0.115,T_3=0.3166,\theta_0=0.2$   &0.00060237 &  0.00095971   &0.000113 &  0.000112   \\ \hline
\end{tabular}
\end{center}
\caption{Check of the equal area law in the $T-\delta S$ plane.}\label{tab3}
\end{table}
It is obvious that $A_L$ equals nearly   $A_R$ for  a fixed   $Q$ and $b$. That is, the equal area law is also valid for the first order phase transition and pseudo phase transition in the entanglement entropy-temperature plane. This result is the same as that of the thermal entropy as well as two point correlation function.

To confirm that $T_c$ is the second order phase transition temperature in the $\delta S-T$ plane, we should explore whether it satisfies a similar relation as Eq.(\ref{c2}).
For different  $b$ and $Q$, the relation between $ \log\mid T-T_c\mid$ and $\log\mid\delta S-\delta S_c\mid  $ are plotted in  Figure \ref{fig14},
in which  $S_c$ is the critical entropy obtained numerically by the equation $T(\delta S)=T_c$. The analytical result of these curves can be fitted as
\begin{equation}
\log\mid T-T_c\mid=\begin{cases}
17.3605 + 2.91954  \log\mid\delta S-\delta S_c\mid,&$for$~b=5, \theta_0=0.2,\\
17.2131 + 2.93553   \log\mid\delta S-\delta S_c\mid, & $for$~b=4, \theta_0=0.2.\\
\end{cases}
\end{equation}
We find for a fixed  $b$ and $Q$, the slope is always about 3, which is consistent with that  of the thermal entropy.  That is, the entanglement entropy  also exhibits a second order phase transition, with phase transition temperature $T_c$, as that of the thermal entropy and  two point correlation function.

\begin{figure}
\centering
\subfigure[$b=5, \theta_0=0.2$]
{\includegraphics[scale=0.75]{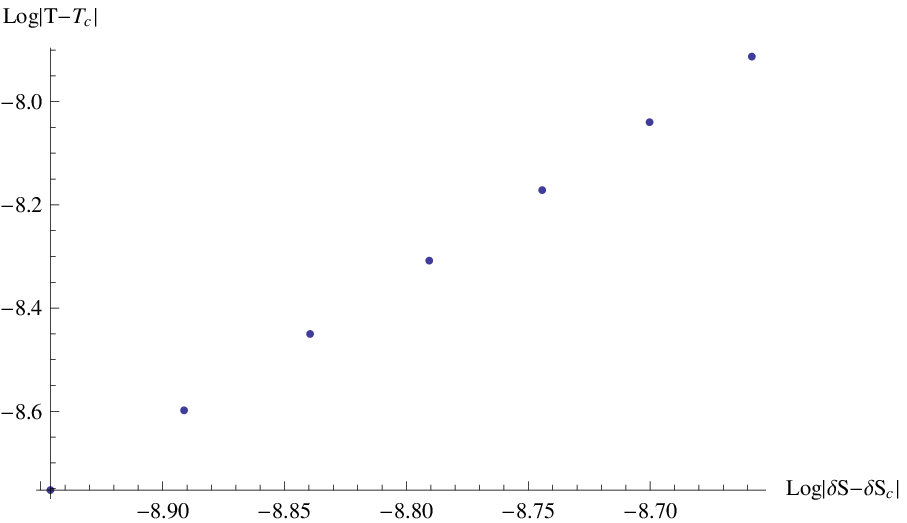}  }
\subfigure[$b=4, \theta_0=0.2$]{
\includegraphics[scale=0.75]{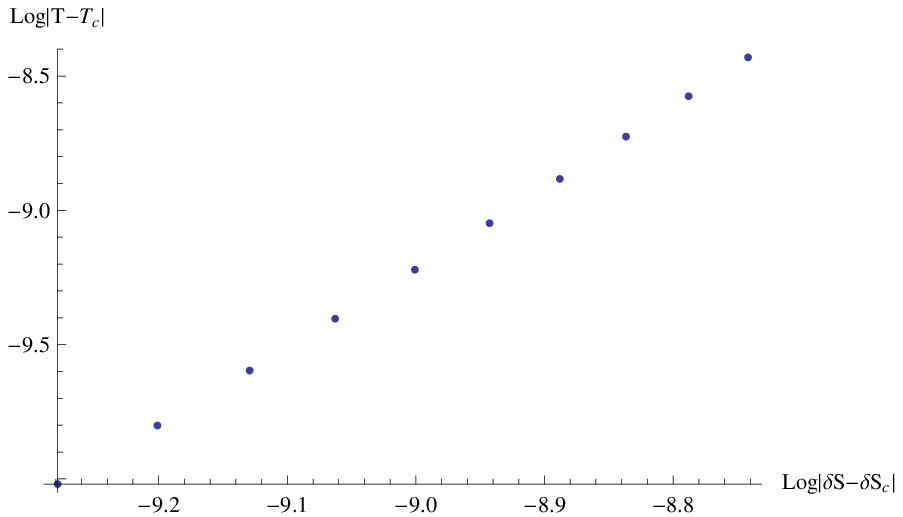}  }
\caption{\small  Relations between $\log\mid T-T_c\mid$ and  $\log\mid\delta S-\delta S_c\mid $ for different $b$ and $\theta_0$.} \label{fig15}
\end{figure}

\section{Concluding remarks}

Investigation on the phase structure of a back hole is of great importance for it reveals whether and how a black hole emerges. Usually, it is realized by studying the relation between some thermodynamic quantities in a fixed ensemble.
In this paper, we found that the phase structure of a back hole also can be probed by the two point correlation function and holographic entanglement entropy, which provides a new strategy to understand the phase structure of the black holes from the viewpoint of holography.

Specially speaking, we first investigated  the  phase structure of the Born-Infeld-anti-de Sitter black holes in the
$T-S$ plane in the fixed charge ensemble.
 We found that the phase structure of the black hole depends on not only the value of $Q$ or $b$,  but also the combination of $bQ$. For the case
$b=5$,  the black hole resembles as the  Reissner-Nordstr\"{o}m-AdS  black hole, and it undergoes the Van der Waals-like phase transition
as its charge satisfies the non-extremal  condition $bQ>0.5$. For the case $b=4$,  besides the Van der Waals-like phase transition, we also observed   a novel phase structure for the case $bQ<0.5$.
With the two point correlation function and holographic entanglement entropy, we further probed the phase structure of the  Born-Infeld-anti-de Sitter black holes and found that both the probes exhibited the same phase structure as that of the thermal entropy for cases $bQ>0.5$ and $bQ<0.5$ regardless of the size of the boundary region. This conclusion was  reinforced by checking the equal area law for the first order phase transition  as well as pseudo phase transition, and calculating the critical exponent of the analogous heat capacity for the second order phase transition.

In previous investigation on the phase structure of the Born-Infeld-anti-de Sitter black holes, authors concentrated mainly on  that of  $bQ>0.5$ \cite{Myung,Lala,Hendibi,Zouni}. In this paper we found that the phase structure for the case  $bQ<0.5$ is also interesting.  As $bQ\ll0.5$, the black hole resembles as  the Schwarzschild AdS black hole as expected, and there is a minimum temperature. We found the larger the charge or the Born-Infeld parameter is, the smaller the minimum temperature will be. That is, both the charge and the Born-Infeld parameter promote the formation of an AdS black hole.
As $bQ$ approaches to 0.5, a new extremal small black hole branch emerges compared with that of the Reissner-Nordstr\"{o}m-AdS  black hole, so that there are two unstable regions and correspondingly two phase transition temperature. The high temperature phase transition was found to be pseudo  for in this case the space time was dominated by the large black hole, which was observed from the
 $F-T$ relation.


\section*{Acknowledgements}
We would like to thank Professor Rong-Gen Cai for his helpful suggestions.
This work  is supported  by the National
 Natural Science Foundation of China (Grant No. 11365008, Grant No. 11205226, Grant No. 11575270), and  Natural Science
 Foundation of  Education Committee of Chongqing (Grant No. KJ1500530).

\end{document}